# The correlation of ~ 27 day period solar activity and daily maximum temperature in continental Australia

Ian Edmonds

Solartran Pty Ltd, 12 Lentara St, Kenmore, Brisbane, Australia Q4069.
ian@solartran.com.au

**Abstract.** We report the first observation of a ~27 day period component in daily maximum temperature recorded at widely spaced locations in Australia. The ~27 day component, extracted by band pass filtering, is correlated with the variation of daily solar radio flux during years close to solar minimum. We demonstrate that the correlation is related to the emergence of regions of solar activity on the Sun separated, temporally, from the emergence of other active regions. In this situation, which occurs only near solar minimum, the observed ~27 day variation of temperature can be in-phase or out-of-phase with the ~27 day variation of solar activity. During solar maximum correlation of temperature and solar activity is much less defined. The amplitude of the ~27 day temperature response to solar activity is large, at times as high as 6$^{o}$C, and much larger than the well documented temperature response to the ~11 year cycle of solar activity. We demonstrate that the ~27 day temperature response is localised to the Australian continent and increases towards the centre of the continent. Several aspects of the ~27 day temperature response are consistent with a ~27 day period solar input to the Earth's continents generating large scale planetary waves.

## 1. Introduction.

Interest in solar activity-climate effects has revived recently, possibly due to the pause in global warming. There was also considerable interest in solar activity-climate effects in the 1970's, Wilcox et al (1974), King et al (1977), Volland (1979). Both the earlier and more recent work in this area was reviewed by Gray et al (2010). The most recognised solar activity-climate effect is the correlation of global surface temperature with the ~11 year solar cycle. This correlation was established by regression of surface temperature against the solar activity proxies such as $F_{10.7}$ radio flux, sunspot number and total solar irradiance, Lean and Rind (2008), Tung et al (2008). The ~11 year cycle of solar forcing, applied to climate models used by the IPPC, contributes a 0.05$^{o}$C amplitude contribution to model predictions of global temperature change, Gray et al (2010), Haigh, (2003). Regression analyses of the ~11 year variation of $F_{10.7}$ flux with climate variables such as temperature, clouds, rainfall, cyclones etc are numerous, Gray et al (2010) and the effect of the ~27 day period solar activity on the upper atmosphere has been well established since 1981, (Ebel et al 1981, Hood and Zhou 1998, Ruzmaikin et al 2007, Forbes et al 2006, Liang et al 2008)). However, aside from the reports in the 1970's there have been few reports of the effects of the ~27 day period solar activity on climate variables in the troposphere. Hood (2003) conducted a cross correlation analysis of temperature at 16 km altitude against various solar proxies over the years 1979-83 and 1989-92 finding a 0.1$^{o}$C, ~27 day period response. Burns et al (2008) found a small (0.9%) correlation of surface pressure against the ~27 day variation of interplanetary magnetic field in Antarctica and the Arctic (opposite phase) during 1995-2005 and interpreted this as due to the effect of solar activity on the global electric circuit. Takahashi et al (2010), compared FFT of $F_{10.7}$ index and OLR over the interval 1980 to 2002 to demonstrate a ~27 day period variation in cloud amount in the Western Pacific but did not propose a mechanism. In this paper we provide evidence of a significant component in Australian surface temperature that is correlated with the ~27 day variation of solar activity. In Section 2 of the paper we describe a band pass filter method of comparing solar activity and daily maximum temperature, $T_{MAX}$, and explain why the method is superior to other methods of detecting association of climate with short term solar activity. In the results section we (1), demonstrate correlation of the filtered components of $F_{10.7}$ flux, sunspot area (SSA) and daily maximum temperature, $T_{MAX}$; (2), estimate the sensitivity of the ~27 day component of $T_{MAX}$ to solar activity; (3), find the spatial distribution of the ~27 day temperature response to solar activity, (4) compare the variation in $T_{MAX}$ directly to images of sunspot activity on the Sun and, (5) use SSA data to extend the study of association of $T_{MAX}$ and solar activity back to 1870. In the discussion section the observations of this paper are related to the Sun induced standing planetary wave mechanism, King et al (1977), Volland (1979).

## 2. Method.

### 2.1 Method and data

We use a method of band pass filtering similar to that used by Lean and Rind (2008) to compare the ~11 year components in surface temperature with solar activity and the method of band pass filtering used by Ma et al (2012) to compare the ~27 day components in $F_2$ region peak electron densities with solar activity. In the present study the dependent variable is daily maximum surface temperature, $T_{MAX}$, in Australia. As with Ma et al (2012) we use daily values of the $F_{10.7}$ index as a proxy for solar EUV activity. The $F_{10.7}$ data was obtained from the National Geophysical Data Centre, ftp://spdf.gsfc.nasa.gov/pub/data/omni/low_res_omni/omni_01_av.dat Sunspot area (SSA) data was obtained from http://solarscience.msfc.nasa.gov/greenwch/daily_area.txt The $T_{MAX}$ data was obtained from the Australian High-Quality Climate Site Network, http://www.bom.gov.au/climate/change/hqsites/ Following Ma et al (2012) the longer term changes in a variable Y(t) are removed by subtracting the 33-day moving average, denoted $S_{33}Y(t)$. In the remainder of this paper the time parameter, (t), is dropped from expressions e.g. Y(t) is referred to as Y where Y is a daily value. The relative short term change, $\Delta Y$, in the variable, Y, is found using

$$\Delta Y = (Y - S_{33}Y)/S_{33}Y \quad (1)$$

Isolation of the ~ 27 day component of the relative variation was obtained by making a Fast Fourier Transform of each of the $\Delta F_{10.7}$, $\Delta$SSA and $\Delta T_{MAX}$ data series. The n Fourier amplitude and phase pairs, $A_n(f_n)$, $\phi_n(f_n)$, in the frequency range 0.031 days$^{-1}$ to 0.045 days$^{-1}$ (period range 32 to 22 days) were then used to synthesize a band pass filtered version of each variable, here denoted $\Delta_{27}F_{10.7}$, $\Delta_{27}$SSA and $\Delta_{27}T_{MAX}$, by summing the n terms, $\Delta_{27}Y_n = A_n\mathrm{Cos}(2\pi f_n t - \phi_n)$. Occasionally the raw data itself is filtered in which case the ~27 day period component is denoted $_{27}Y$.

## 2.2 Method efficiency.

Here we outline the reason for using a band pass filter method to detect association between solar activity and temperature rather than the more commonly used averaging methods of spectral, correlation or epoch superposition analysis. In the variation of $F_{10.7}$ solar radio flux, Figure 1, the presence of a ~11 year solar cycle and a ~27 day cycle associated with solar rotation is evident. It is clear that the amplitudes of the two cycles are comparable; about 50 solar flux units (sfu, $10^{-22}$ $Wm^{-2}Hz^{-1}$) for the ~11 year period cycle and about 50 sfu near solar maximum and about 10 sfu near solar minimum for the ~27 day period cycle.

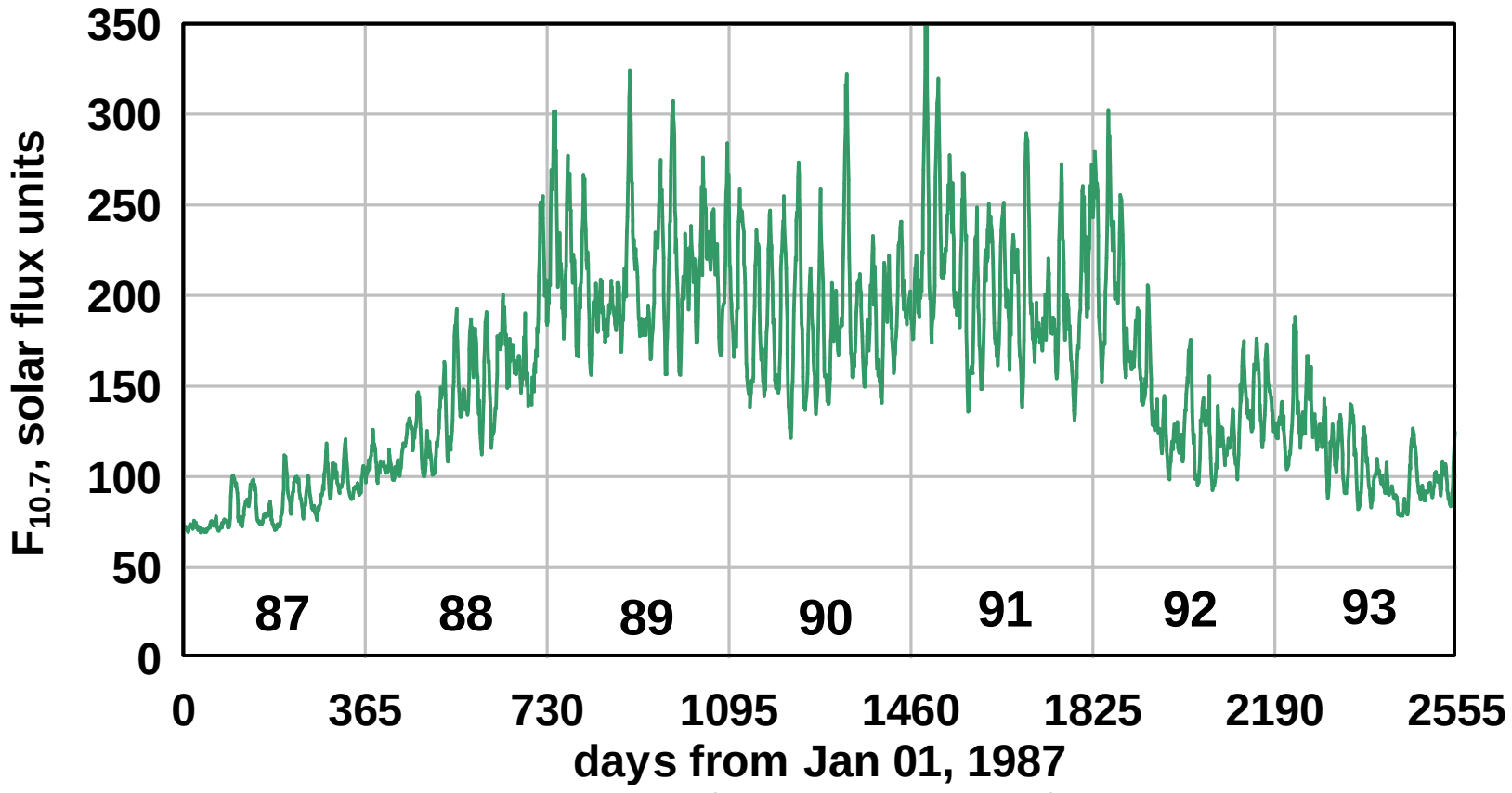

Figure 1. Approximately one ~11 year cycle of $F_{10.7}$ solar radio flux indicating the relative amplitudes of the ~11 year cycle and the ~27 day cycle.

One might expect, therefore, that when a frequency spectrum is obtained over a $F_{10.7}$ index record that peaks of comparable amplitude, about 50 sfu at ~11 years and about 30 sfu at ~27 days, would be obtained. The FFT for the interval 1987 to 2012 is shown in Figure 2. The ~11 year peak is, as expected, about 50 sfu but the ~27 day peak is about 3 sfu. It is clear that the efficiency of detecting the ~27 day variation is low. The reason for this is now examined.

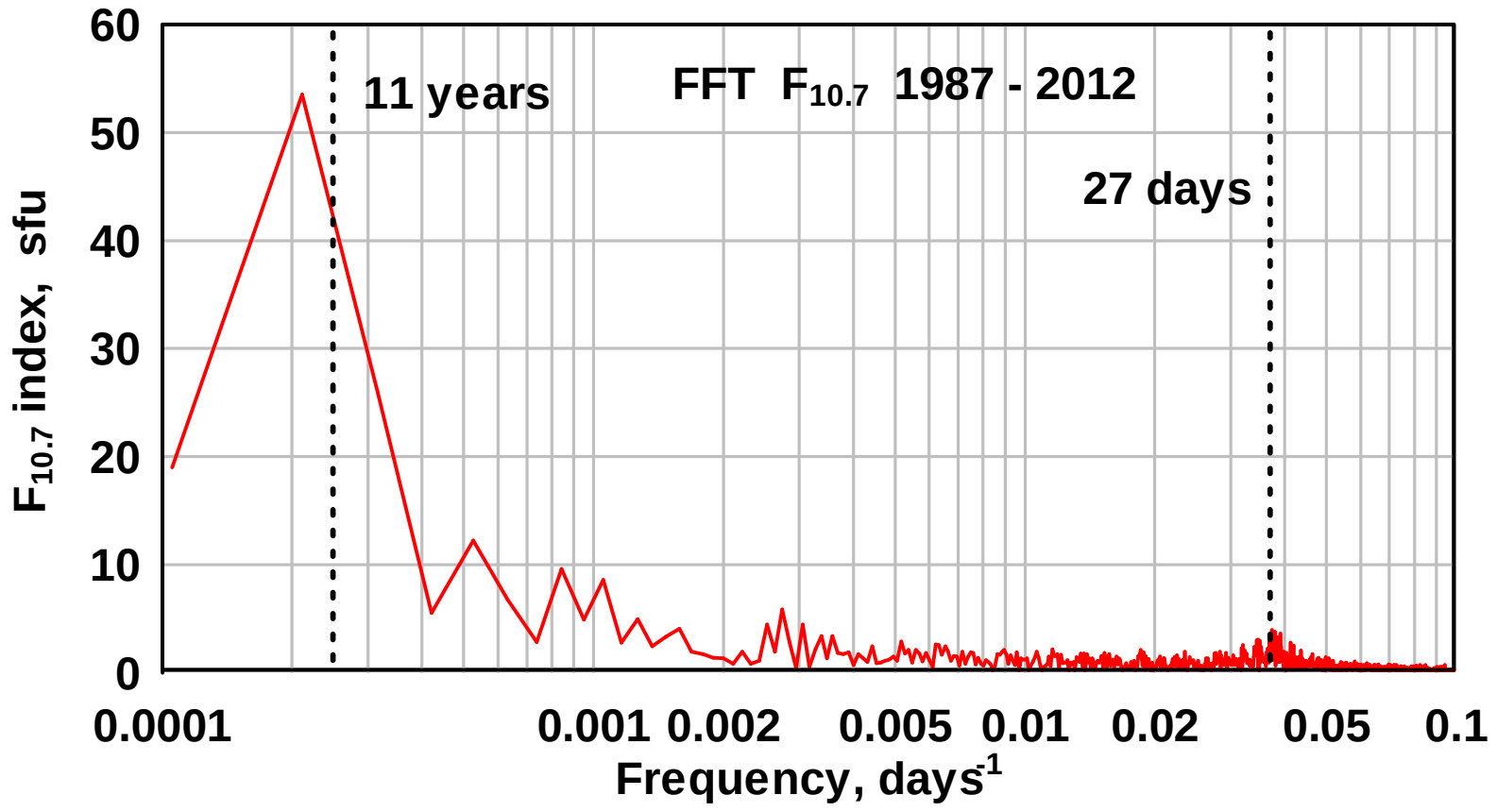

Figure 2. FFT of the daily F10.7 solar radio flux in the years 1987 to 2012.

Figure 3 shows the $\Delta_{27}F_{10.7}$ for 1975 and the years 2010 and 2011; all years near solar minimum. In each graph it is possible to follow, in time, the effect on $F_{10.7}$ flux of the emergence and evolution of individual sunspot groups or active regions on the Sun; in 1975 three sunspot groups evolved; in 2010-2011 five sunspot groups evolved. During solar minimum sunspot group evolutions (SGE) are relatively infrequent and the modest overlapping of one evolution with another is evident in Figure 3. At solar maximum, when several sunspot groups are evolving simultaneously, the situation is more complex. The time axis in each graph of Figure 3 is spaced at 27 day intervals indicative of solar rotation. It is evident that the phase lag relative to solar rotation is different for each SGE. For example, in 1975, the phase of the variation during the first evolution is shifted by about $\pi$ radians relative to the phase of the variation during the third evolution.

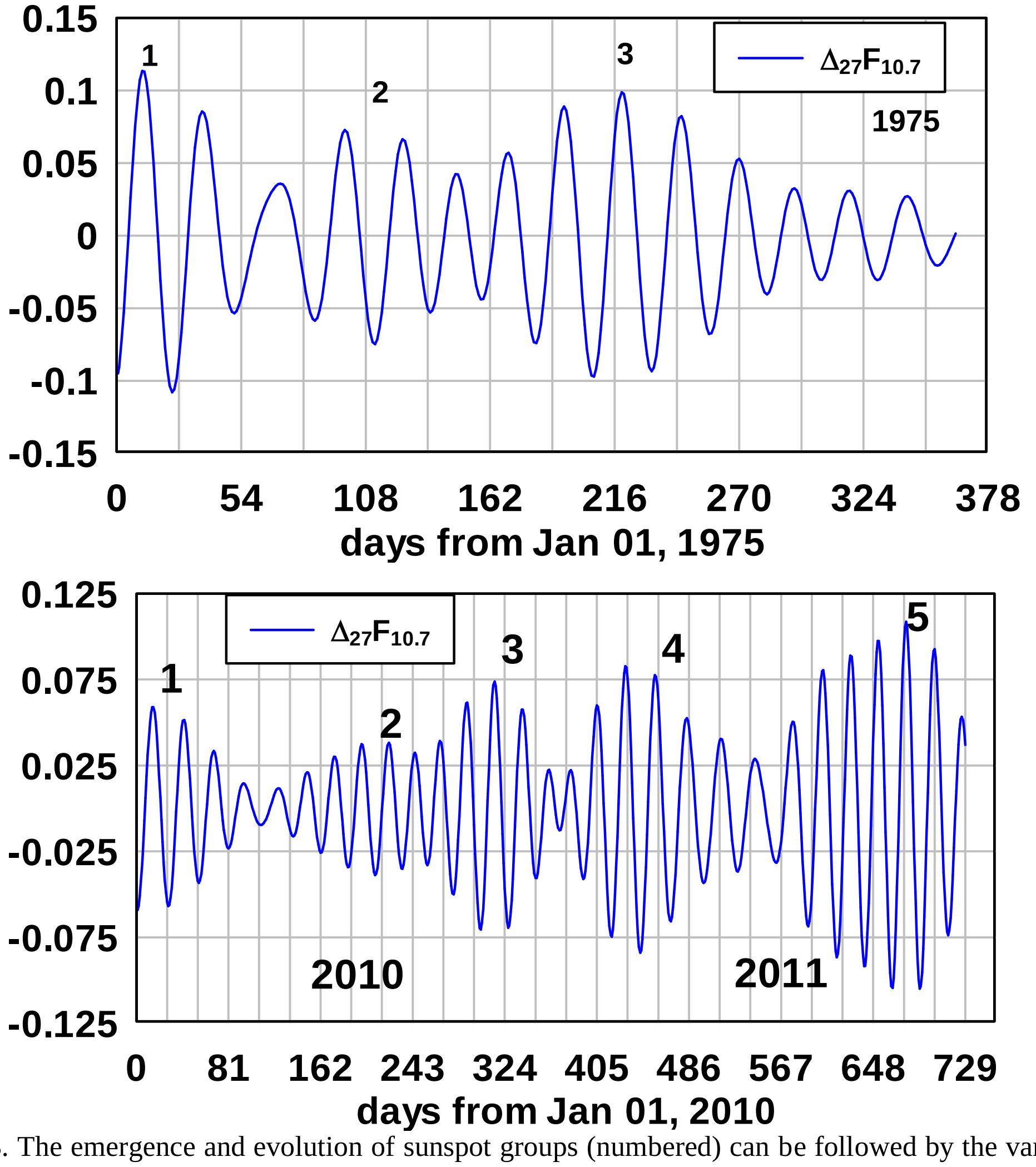

Figure 3. The emergence and evolution of sunspot groups (numbered) can be followed by the variations in $F_{10.7}$ solar radio flux.

The different phase lags occur because individual sunspot groups emerge at random longitudes on the Sun. The component of a Fourier spectrum at frequency, f, is obtained from the integration of product terms between the variable of interest and sinusoid and co-sinusoid pairs of frequency f. As the phase lag of the $F_{10.7}$ variable varies, with different SGE, over the range between 0 and $2\pi$, the product terms will vary between positive and negative. Thus the integral of the product term accumulates, not as N, where N is the number of SGE, but as $N^{1/2}$ as in a random walk or Markov process. As the length of the record over which the spectrum is obtained increases the efficiency of detecting a component at ~27 days period decreases as $1/N^{1/2}$. For example, the average duration of a SGE is about three solar rotations or 81 days. Therefore, during the 9125 day interval of 1987 to 2012 about N = 113 SGE are expected to occur. The efficiency is $1/N^{1/2} = 0.094$ and the expected peak height at ~27 day period is about 30 x 0.094 = 2.8 sfu, about the level observed in Figure 3. This decrease in the efficiency of spectral analysis obviously applies to any record of solar activity that depends directly on SGE such as $F_{10.7}$, SSN, and SSA records. However, it also applies to any variable that is connected with the solar activity. Therefore the efficient use of spectral analysis in detecting ~27 day period association between solar activity and climate variables is limited to record lengths encompassing one, or at the most a few SGE. For example, wavelet analysis would be efficient provided the wavelet duration is comparable with the average duration of a SGE, typically three solar rotations or 81 days.

Correlation and epoch superposition studies that compare the solar activity variation itself with climate variation, e.g. King et al (1977), Hood (2003), Burns et al (2008), would, at first sight, seem to be unaffected by the above considerations. However, in this paper we will show that there is a variation in phase lag between solar activity and temperature for different SGE. The cause of this variable phase lag must be different from the effect discussed above and is discussed later. However, the effect on the efficiency of averaging methods such as correlation and epoch superposition is essentially the same as the effect on frequency analysis i.e. the efficiency of detecting association decreases as the length of the record or the number of SGE encountered increases. For this reason, in this paper, we detect association between solar activity and temperature by direct comparison, over time intervals of a few SGE, of band pass filtered or unfiltered records of solar activity and surface temperature. Unfiltered records of solar activity are used where appropriate to: (1), minimise the processing of data; (2), retain detail useful in confirming association, and (3), to avoid confusing the band pass filter response to a spike in activity (a sinc function) with the band pass filter response to the evolution of a sunspot group over several solar rotations (similar to a sinc function).

## 3. Results.

### 3.1 Comparison of the $\Delta F_{10.7}$ variation with the $\Delta_{27}T_{MAX}$ variation for Melbourne.

The daily $T_{MAX}$ record for Melbourne Regional station (38S, 145E) is used to assess the correlation of $\Delta F_{10.7}$ and $\Delta_{27}T_{MAX}$ in the interval 1969 – 1986 examined by Ma et al (2012). It is appropriate to use the $\Delta F_{10.7}$ variation rather than the $\Delta_{27}F_{10.7}$ variation for comparison as the $\Delta F_{10.7}$ variation is predominantly a ~ 27 day period variation and band pass filtering is largely superfluous to isolate the ~ 27 day variation. On the other hand the $\Delta T_{MAX}$ variation is a combination of a ~ 27 day variation and other very significant short term variations so band pass filtering is usually essential to resolve the ~ 27 day variation. Figure 4 compares $\Delta F_{10.7}$ (dotted curves) and $\Delta_{27}T_{MAX}$ for 1969, 1975 and 1986.

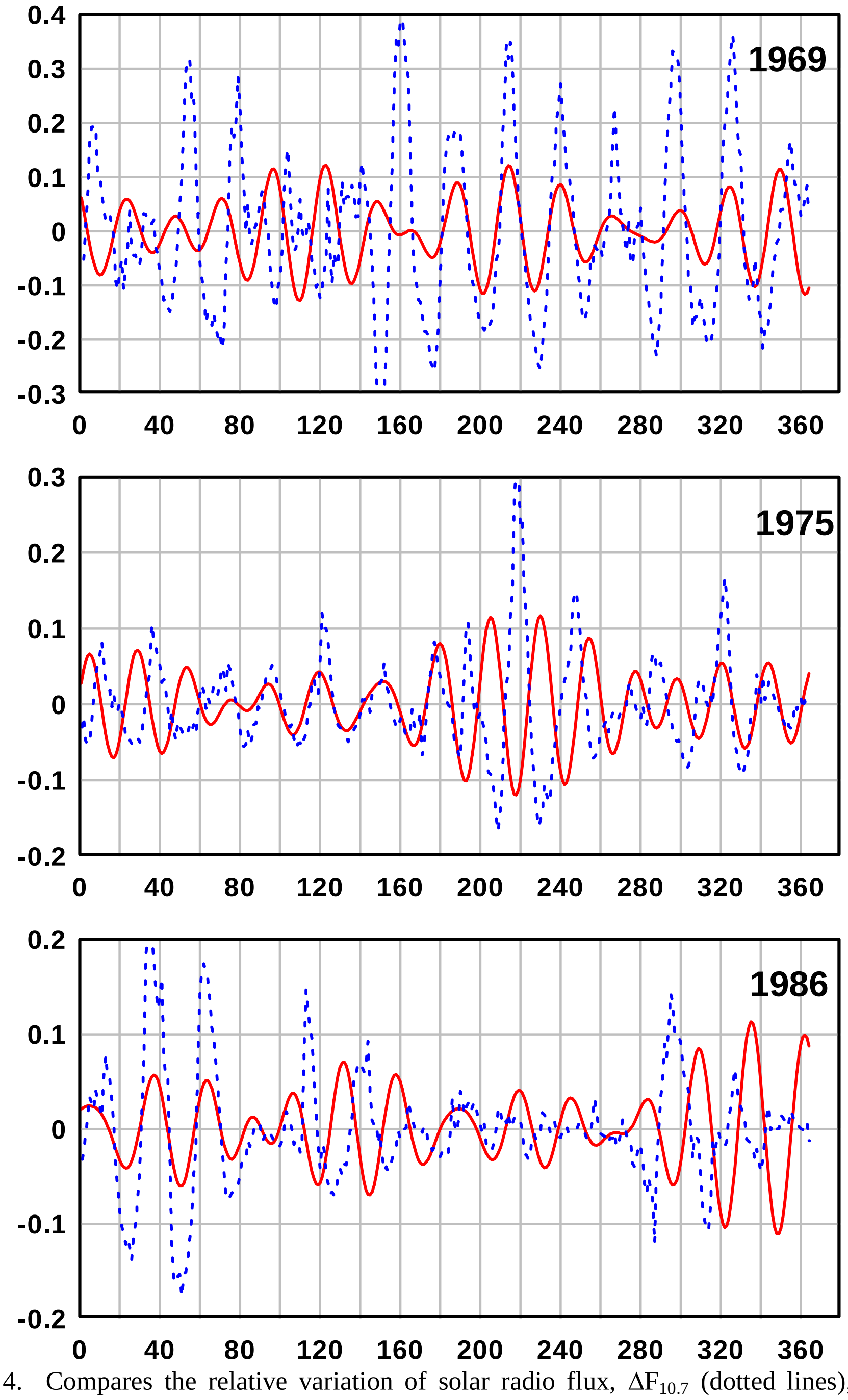

Figure 4. Compares the relative variation of solar radio flux, $\Delta F_{10.7}$ (dotted lines), and the band pass filtered, relative variation of daily maximum temperature, $\Delta_{27}T_{MAX}$ (full lines), for Melbourne during years, 1969, 1975 and 1986.

In 1969 $\Delta F_{10.7}$ and $\Delta_{27}T_{MAX}$ show weak and negative correlation for the first half of the year and strong positive correlation for the second half of the year. In 1975 and 1986 the correlation of the variables alternates between positive and negative as the $\Delta F_{10.7}$ variation passes through periods of stronger activity associated with SGE. The records for 1975 and 1986 show amplitude correlation with the intervals of stronger $\Delta_{27}T_{MAX}$ variation clearly associated with the intervals of stronger $\Delta F_{10.7}$ variation. It is evident that, during 1975 and 1986, the intervals of stronger $\Delta F_{10.7}$ activity correspond to intervals of SGE and that the average SGE lifetime is about three or four solar rotations. The change from mainly in-phase variation (positive correlation) of $\Delta F_{10.7}$ and $\Delta_{27}T_{MAX}$ to mainly out-of-phase variation (negative correlation) of $\Delta F_{10.7}$ and $\Delta_{27}T_{MAX}$ is also evident in other years, e.g. 1977 and 1985, provided one SGE does not overlap with the previous or the next SGE. Other examples, shown later in this paper, occur in 1927, 1928 and 2011.

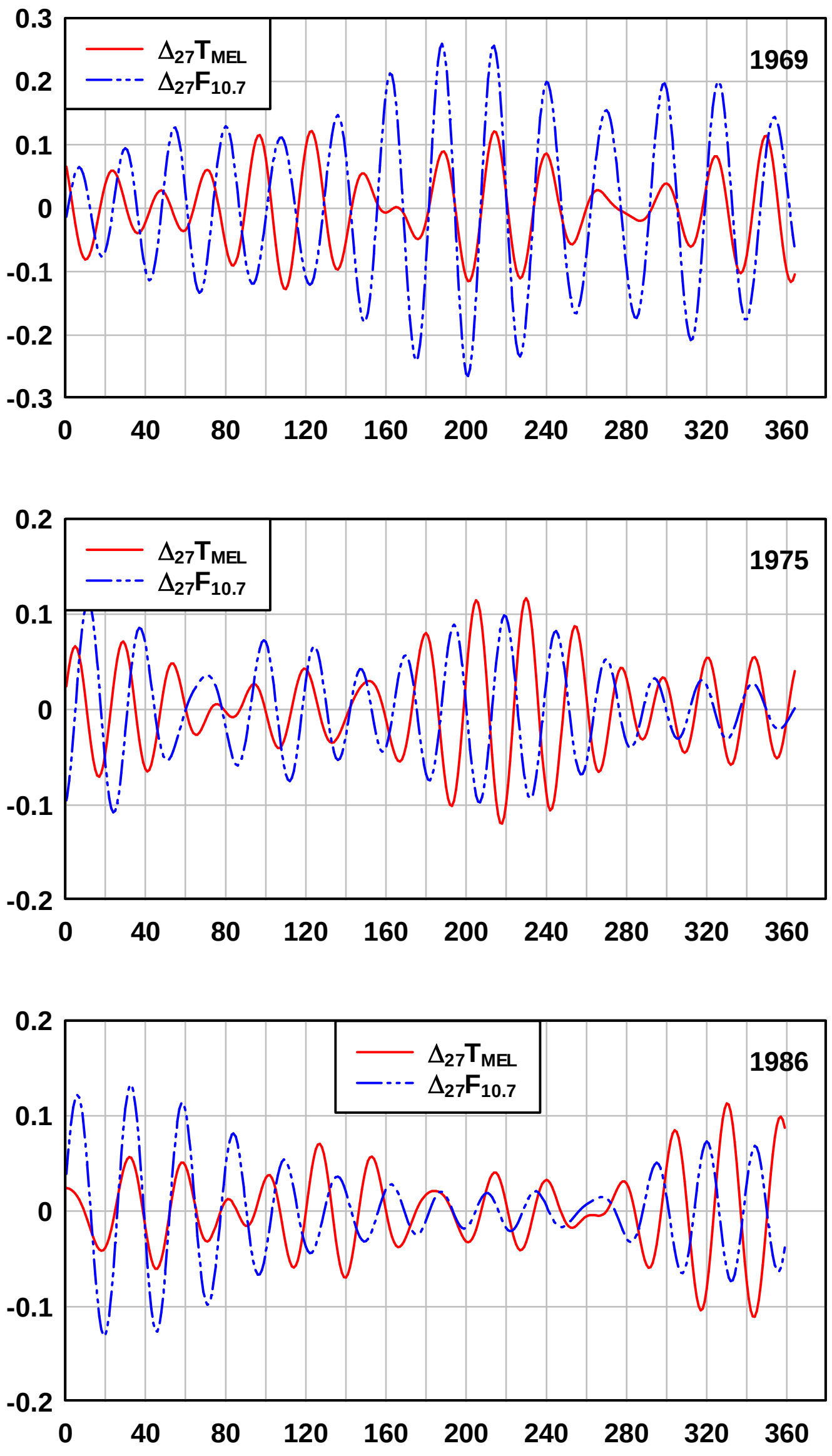

Figure 5. Band pass filtered components of F10.7 solar radio flux and daily maximum temperature in Melbourne for 1969, 1975 and 1986.

When one SGE overlaps with the emergence of the next or the previous SGE, as appears to be the case, for example, in the first half of 1969, the phase and amplitude relationship between $\Delta F_{10.7}$ and $\Delta_{27}T_{MAX}$ is less defined. Comparison of the band pass filtered components of both variables, Figure 5, allows better definition of the phase lag associated with each SGE. Variation near in-phase or near out-of-phase predominates, and near quadrature phase can occasionally be discerned e.g. the first part of 1975. In the transition interval between one SGE and the next the phase lag is changing and indefinite. Observations of $\Delta F_{10.7}$ and $\Delta_{27}T_{MAX}$ that illustrate the situation, near solar maximum, when SGE overlap are shown in Figure 6. A noticeable feature in Figure 6 is that both the amplitude of $\Delta F_{10.7}$ and amplitude of $\Delta_{27}T_{MAX}$ are relatively constant during both 1974 and 1980. This can be compared with the situation near solar minimum, 1975 and 1986 in Figures 4 and 5, when the amplitudes of $\Delta F_{10.7}$ and $\Delta_{27}T_{MAX}$ increase and decrease in step over the duration of each SGE. Near solar maximum, the variation of phase lag, except in short intervals when the amplitude of $\Delta F_{10.7}$ is high, is complex. As a result, during 1974 and 1980, it is difficult to identify consistent amplitude or phase connection between the two variables or with any SGE.

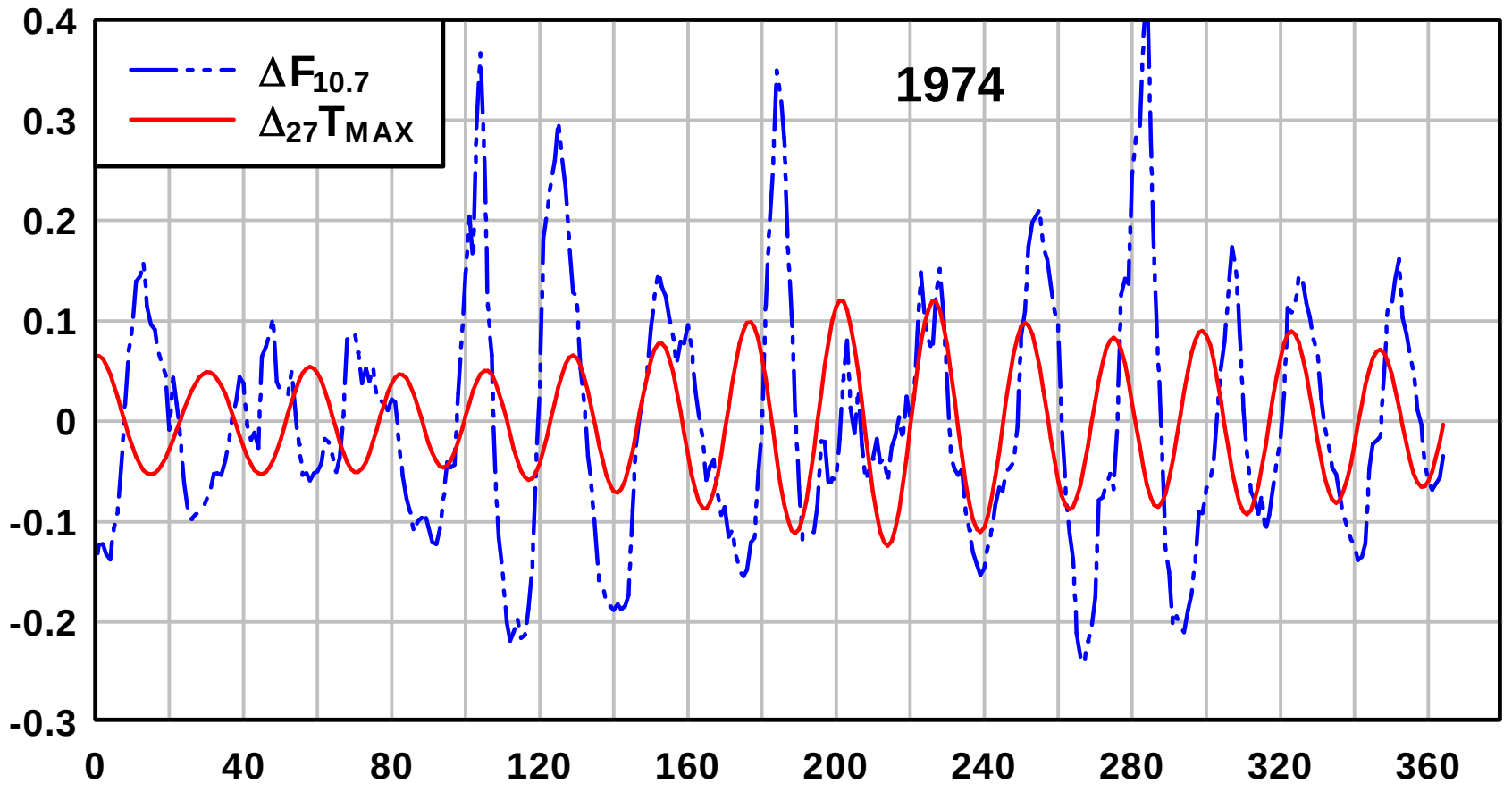

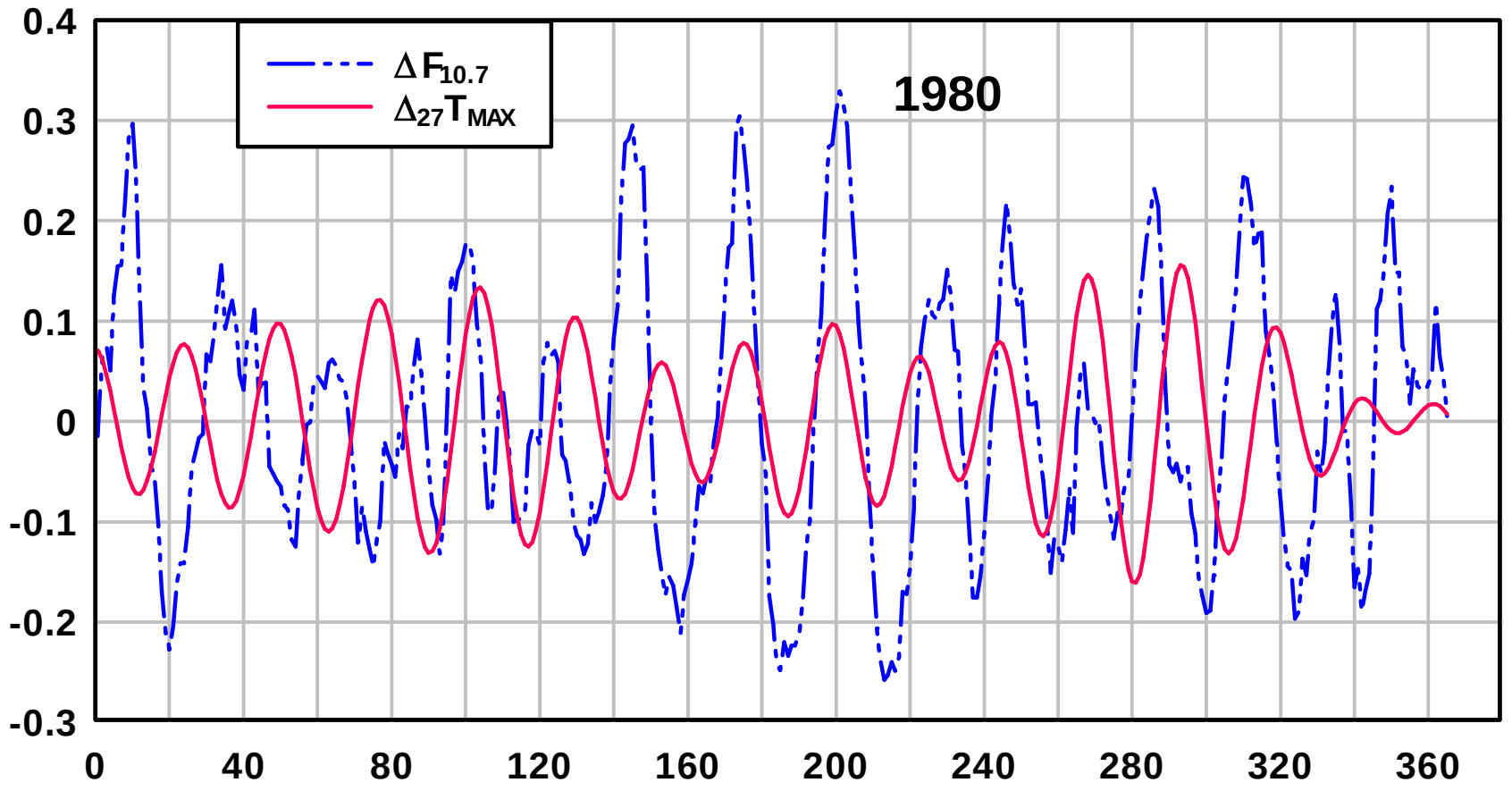

Figure 6. Compares the relative variation of solar radio flux, $\Delta F_{10.7}$ (dot-dash lines), and the band pass filtered, relative variation of daily maximum temperature, $\Delta_{27}T_{MAX}$ (full lines), for Melbourne during years, 1974, and 1980.

### 3.2 The contribution of the ~ 27 day variation in $F_{10.7}$ to the variation in $T_{MAX}$.

A variable Y can be regenerated from the relative change in the variable, $\Delta Y$, by inverting equation 1. Thus

$$Y = S_{33}Y^{*}\Delta Y + S_{33}Y \tag{2}$$

The product $S_{33}Y^{*}\Delta Y$ regenerates the actual amplitude of the Y variation and the addition of $S_{33}Y$ restores the actual baseline. Any separate contribution to $\Delta Y$ can be similarly regenerated so that the ~ 27 day period contribution to $T_{MAX}$ can be readily compared with the overall $T_{MAX}$ variation by use of (2) as in Figure 7.

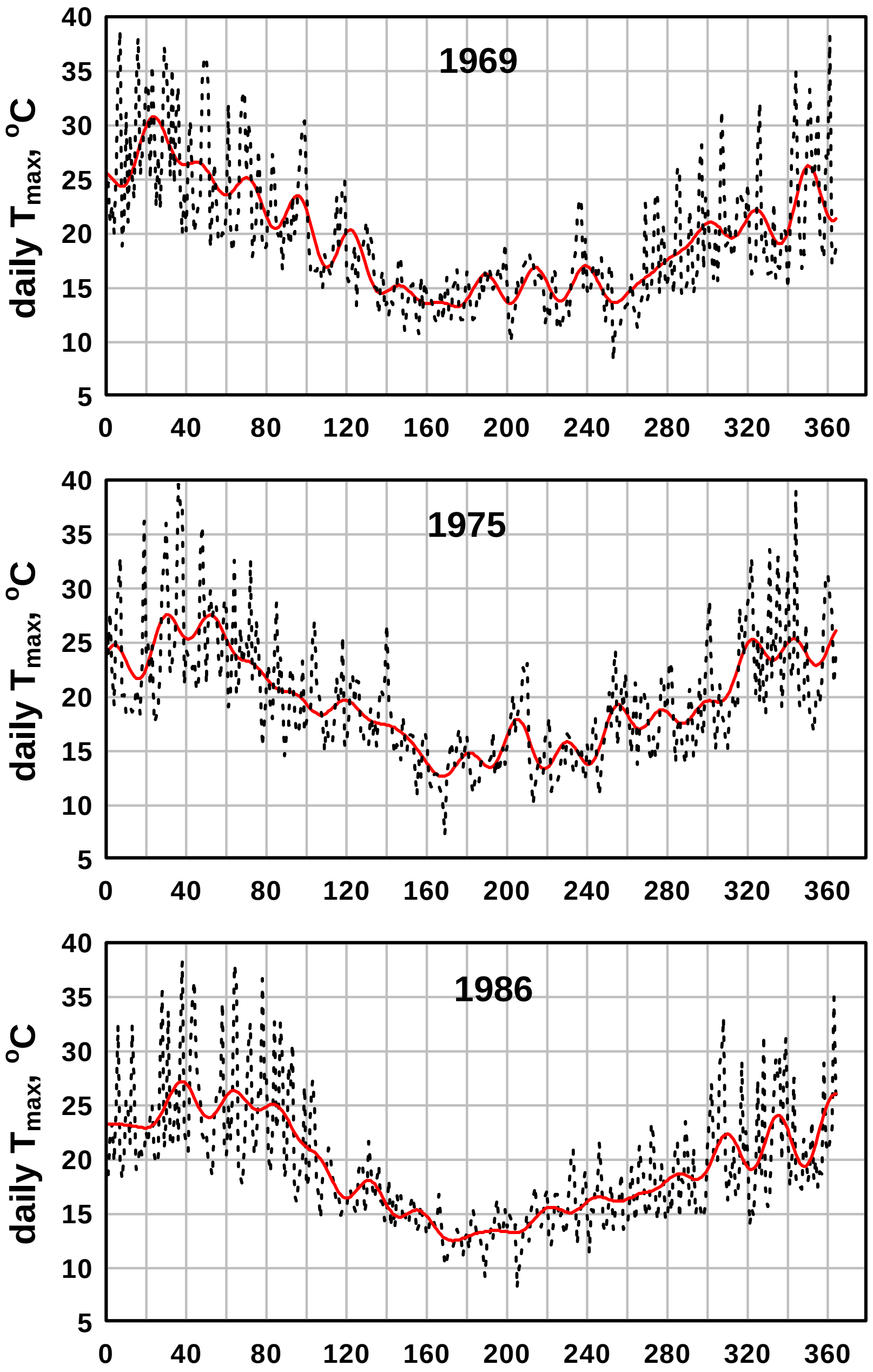

Figure 7. Compares the variation of daily maximum temperature, (dotted lines) with the regenerated version of the band pass filtered ~27 day component of daily maximum temperature (full lines) for 1969, 1975 and 1986 in Melbourne.

Figure 7 shows the regenerated ~ 27 day variation of $T_{MAX}$, known from Figure 5 to be correlated with daily $\Delta F_{10.7}$ values, superimposed on the daily $T_{MAX}$ variation in Melbourne for the years 1969, 1975 and 1986. The graphs also show that the change of the average level of $T_{MAX}$ between summer and winter in Melbourne is about 12$^{o}$C. The variation in the ~27 day period component of $T_{MAX}$ can be as high as 7$^{o}$C peak to peak, see for example the last days of 1969 or 1986. Thus the ~ 27 day period solar UV related component can have a very significant effect on day to day temperature in Melbourne to the extent that peak or minimum daily values of $T_{MAX}$ in any given period can often be associated with peaks or minima in $\Delta_{27}T_{MAX}$. Figures 4 and 5 show that in the second half of 1969 and the first part of 1986 peaks in $\Delta_{27}T_{MAX}$ are consistently associated with peaks in $\Delta F_{10.7}$. However, around day 220 in 1975 and at the end of 1986 peaks in $\Delta_{27}$Tmax coincide with minima in $\Delta F_{10.7}$. So we have the interesting observation that peaks in solar activity may result in peaks or in minima of temperature. The reason why the ~27 day component of temperature can vary either in-phase or out-of-phase with solar activity during a SGE is one of the outstanding problems arising from this work and is discussed later. However this observation goes some way to explaining why this very significant ~27 day period variation in surface temperature has not been previously reported. While the correlation may be strongly positive or strongly negative during a SGE, e.g. in the second half of 1969 and in the middle of 1975, the correlation obtained over a long record encompassing many SGE will, as discussed in section 2, tend to zero. For example, the correlation coefficient of $\Delta_{27}F_{10.7}$ and $\Delta_{27}T_{MAX}$ obtained over the entire 1969 to 1986 record is 0.07.

### 3.3 The sensitivity of surface temperature to solar radio flux variation.

To facilitate a very preliminary estimate of the sensitivity of $T_{MAX}$ to $F_{10.7}$ flux we reproduce the observations during the second half of 1969, when $\Delta_{27}T_{MAX}$ varied in phase with $\Delta F_{10.7}$, in Figure 8.

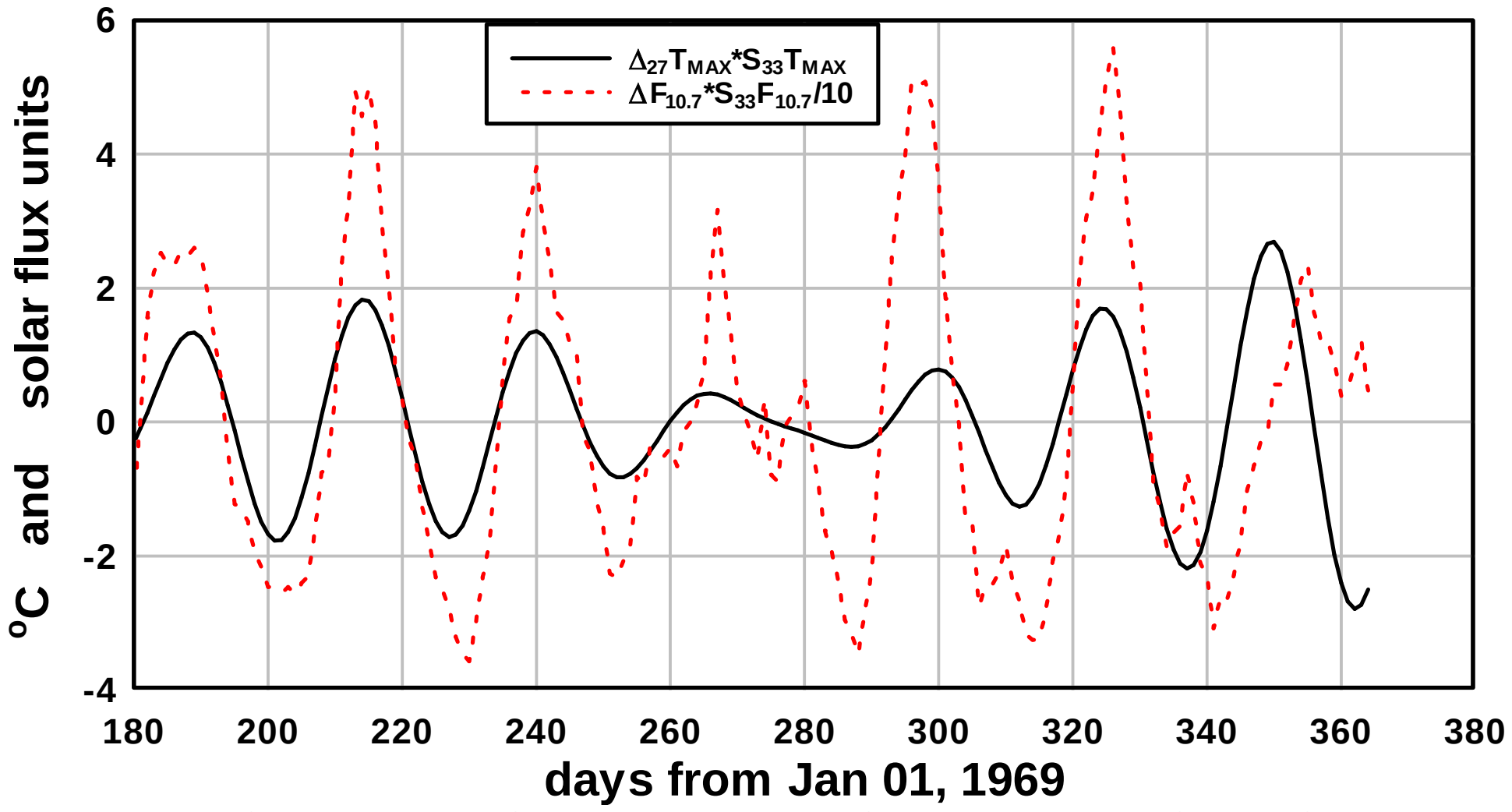

Figure 8. Comparison of the variation of the $F_{10.7}$ solar radio flux, reduced by a factor of 10, (dotted line), and the ~27 day component of $T_{MAX}$ (full line) during the second half of 1969.

The variation of $T_{MAX}$ shown in Figure 8 has been regenerated from $\Delta_{27}T_{MAX}$ values to the actual ~27 day period variation of $T_{MAX}$ by multiplication of the $\Delta_{27}T_{MAX}$ values by the corresponding $S_{33}T_{MAX}$ values. Similarly the $\Delta F_{10.7}$ relative variation has been converted to the actual $F_{10.7}$ variation and this variation has been reduced by a factor of 10 in Figure 8 to facilitate comparison with the temperature variation. The standard deviations of the variations shown in Figure 8 are, respectively 1.14 K and 2.27 solar flux units with the latter value corresponding to 22.7 solar flux units for the standard deviation of the $F_{10.7}$ variable. Thus the average sensitivity of ~27 day $T_{MAX}$ variation to the $F_{10.7}$ variation during the seven solar rotations in the period illustrated is 1.14/22.7 = 0.05 K/solar flux unit. Typically, the ratio of total solar irradiance (TSI) change to $F_{10.7}$ change is approximately 0.0125 $Wm^{-2}$/solar flux unit, Svaalgard (2013). Thus the sensitivity of ~ 27 day $T_{MAX}$ variation to the TSI variation in 1969 is about 0.05/0.0125 = 4 $K/Wm^{-2}$. This sensitivity is many times higher than the sensitivity of the eleven year cycle of global temperature to the eleven year cycle of TSI, 0.12 – 0.17 $K/Wm^{-2}$, reported by Tung et al (2008). This suggests the possibility of, either a very much higher sensitivity of global surface temperature to the ~27 day solar cycle, or, that the phenomenon reported here is localised to the Australian continent. This latter aspect is considered in the next section.

**3.4 Latitude and longitude dependence of the temperature response to solar activity.**
Spatial correlation requires that variables recorded at widely spaced locations vary in the same manner at the same time. To identify how the $\Delta_{27}T_{MAX}$ response varies spatially it is useful to select a year during which a SGE emerges from a low activity background and track, spatially, on Earth, the $\Delta_{27}T_{MAX}$ response to forcing by effects associated with that SGE. During 1975 a well defined SGE occurs between day 160 and day 280 and is associated with a large central peak of $\Delta F_{10.7}$ activity at 218 days that can serve as a useful reference, see Figure 4. Smaller SGE occur at the beginning and end of 1975 but do not overlap significantly with the central SGE. Figure 9 shows the variation of the regenerated ~27 day $T_{MAX}$ response for stations that range, roughly evenly spaced, from latitude 54°S at Macquarie Island to latitude 12°S at Darwin. A scaled-to-fit version of $\Delta F_{10.7}$ is shown for reference. Clearly the temperature response to $\Delta F_{10.7}$ increases towards the centre of the Australian continent with the temperature response at Alice Springs (24°S) being about twice the response at Melbourne (38°S) and much greater than the response at Macquarie Island to the south of the continent and the response at Darwin near the Northern edge. However, there does appear to be a symmetric, if much reduced response, at Macquarie Island and at Darwin, to the central SGE. There also appears to be a shift of the temperature response to slightly later times as the latitude of observation moves up through central Australia. Similar changes are also seen in the temperature response to the smaller SGE that occurs at the start of the year.

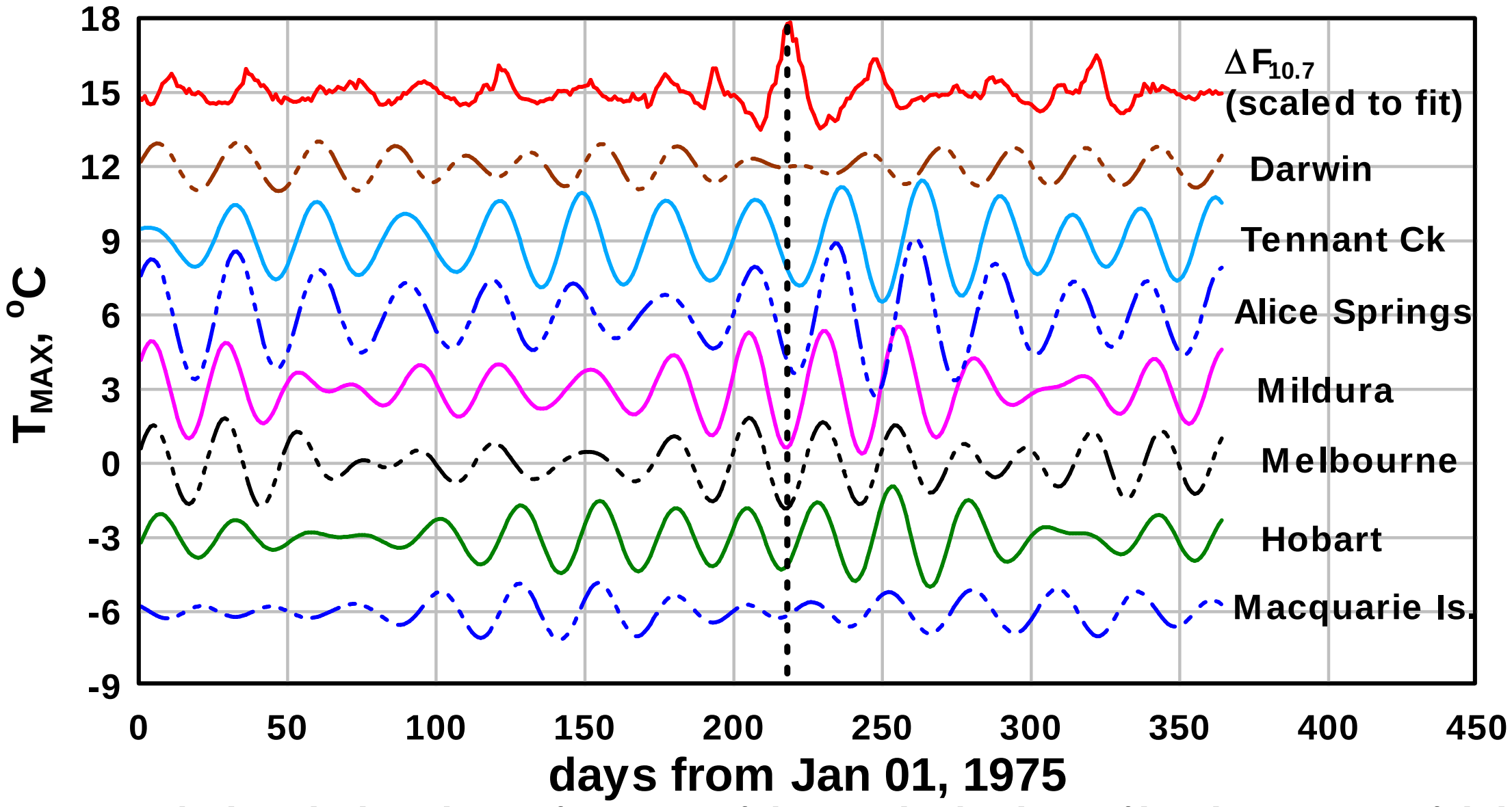

Figure 9. The latitude dependence of response of the ~27 day band pass filtered component of daily maximum temperature, $T_{MAX}$, to the isolated emergence of a region of solar activity as indicated by the $F_{10.7}$ activity centred on day 218 of 1975 (dotted vertical marker).

Figure 10 shows the variation of the regenerated ~ 27 day $T_{MAX}$ response for station locations, roughly evenly spaced, over a longitude range across southern Australia that varies from 115°E at Perth to 168°E at Norfolk Island. The temperature response is in anti-phase with $\Delta_{27}F_{10.7}$ and is significant over the southern, central, land mass of Australia, between Ceduna and Sydney. However, the response at Perth appears uncorrelated to the $\Delta_{27}F_{10.7}$ variation. The response at Norfolk Island, 1400 km off the East coast of Australia, shows a very small, but apparently symmetric, response to the central $\Delta F_{10.7}$ variation. It is remarkable that the response at Ceduna, Melbourne and Sydney are all exactly out of phase with the central $\Delta F_{10.7}$ variation despite the fact that Sydney is about 1200 km (18° of Longitude) East of Ceduna. The observations in Figure 9 and 10 indicate that the ~ 27 day temperature response is spatially correlated over most of the Australian continent. However the response does not extend in latitude to Darwin at 12°S or to Macquarie Is at 54°S. Similarly the response does not extend in longitude to Perth at 115°E or to Norfolk Is at 168°E. Thus a coherent response is localised between 12°S and 54°S and between 115°E and 168°E. A temperature response confined to the Australian continent might be expected as the temperature response to solar exposure is much greater over land than over sea. However, the observed spatial distribution of temperature response is also consistent with an anti-node of a standing wave in the atmosphere induced by ~27 day period heating of the continent, Volland (1979), King et al (1977), as discussed later.

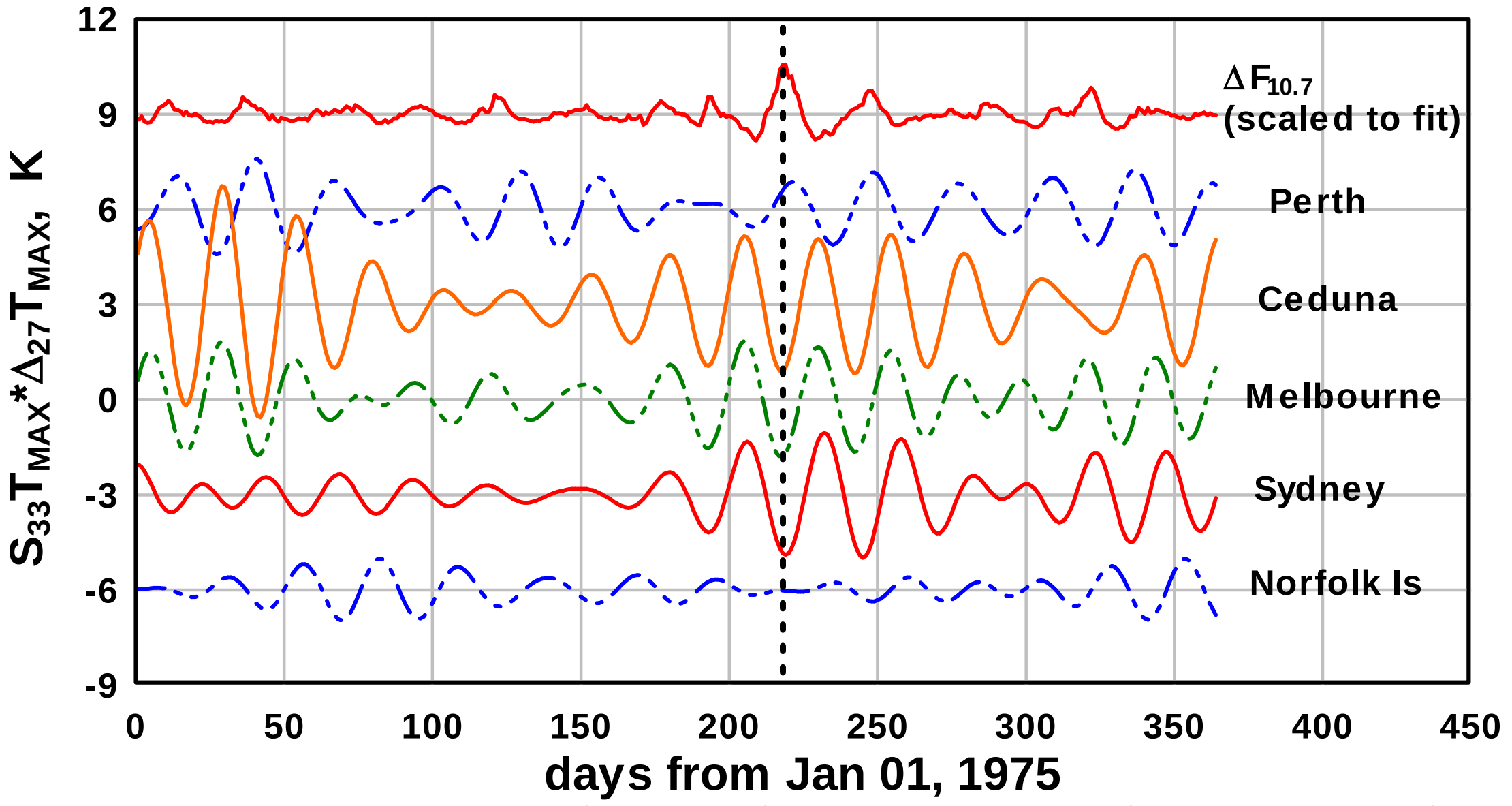

Figure 10. The longitude dependence of response of the ~27 day band pass filtered component of daily maximum temperature, $T_{MAX}$, to the isolated emergence of a region of solar activity as indicated by $F_{10.7}$ activity centred on day 218 of 1975 (dotted vertical marker).

### 3.5 Correlation of $\Delta_{27}T_{MAX}$ with surface activity on the Sun.

Much of the discussion in this paper centres on solar and climate variation during SGE. Data on SGE can be obtained from daily sunspot number or, in more detail, from daily sunspot area records. However, it is useful to directly observe the actuality of a sun group evolution. From June 2010 to the present daily high resolution images of solar activity on the surface of the Sun are available for download from the Solar Dynamics Observatory site at http://sdo.gsfc.nasa.gov/data/aiahmi/index.php . The solar images allow the visual correlation of the $\Delta_{27}T_{MAX}$ variation with the daily variation of solar activity on the surface of the Sun. During 2011 an SGE occurred during the second half of 2011. Figure 11 shows the variation of $\Delta_{27}F_{10.7}$ and the variation of $\Delta_{27}T_{MAX}$ for Melbourne Regional station during 2011.

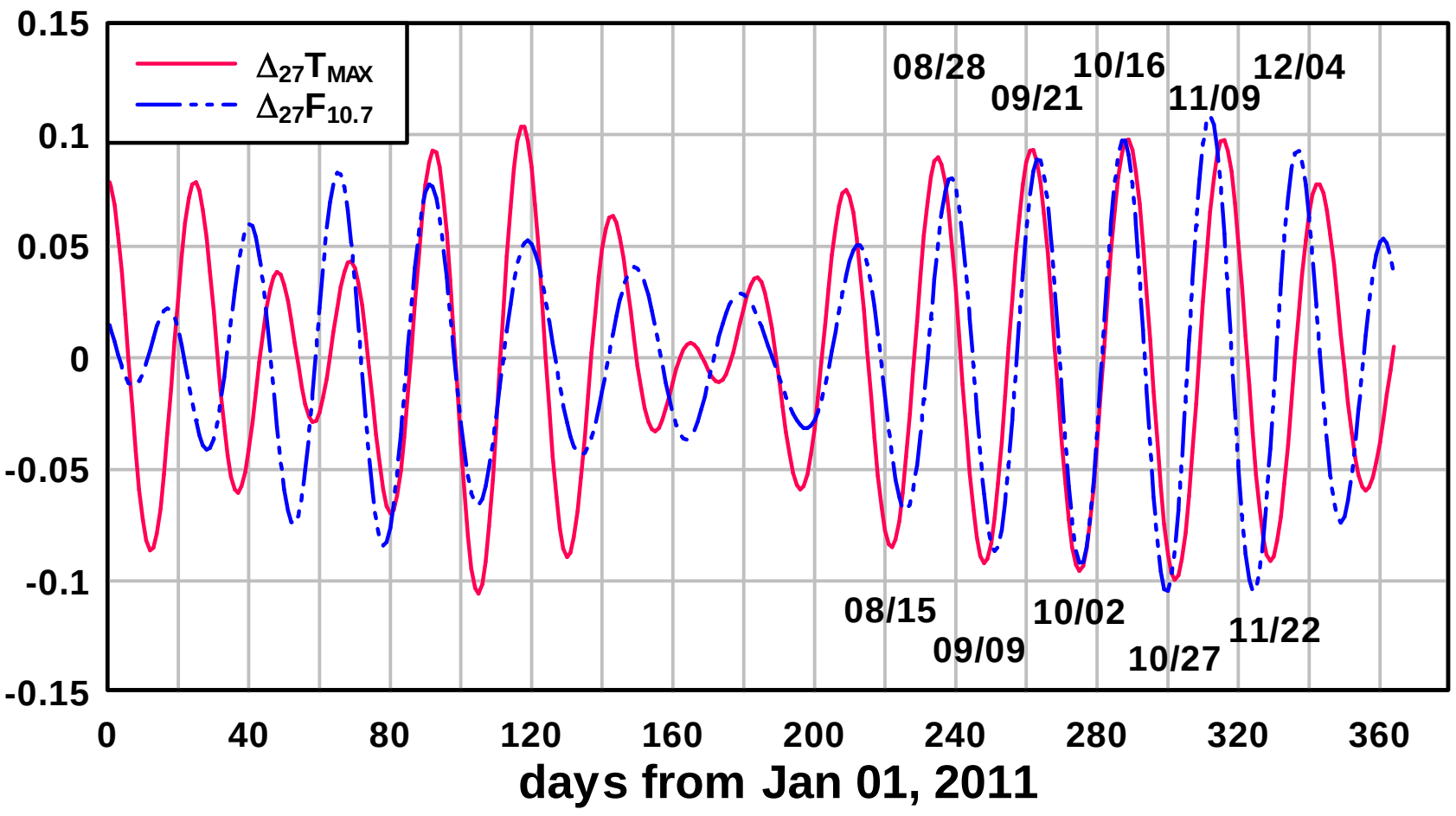

Figure 11. Compares the band pass filtered relative component of daily maximum temperature, $\Delta_{27}T_{MAX}$ (full line), with the band pass filtered relative component of daily solar radio flux, $\Delta_{27}F_{10.7}$ (dot-dash line) during 2011. The dates of maxima and minima of $\Delta_{27}F_{10.7}$ are provided to facilitate comparison with images of the emergence of a solar active region on the Sun during the second half of 2011, (Figure 12).

It is evident that the ~27 day components of the solar activity and the surface temperature are varying approximately in phase over most of 2011. The dates of a sequence of minima and maxima of $\Delta_{27}F_{10.7}$ that occur in the second half of 2011 are also indicated in Figure 11. The sequence of solar images obtained at these dates is shown in Figure 12.

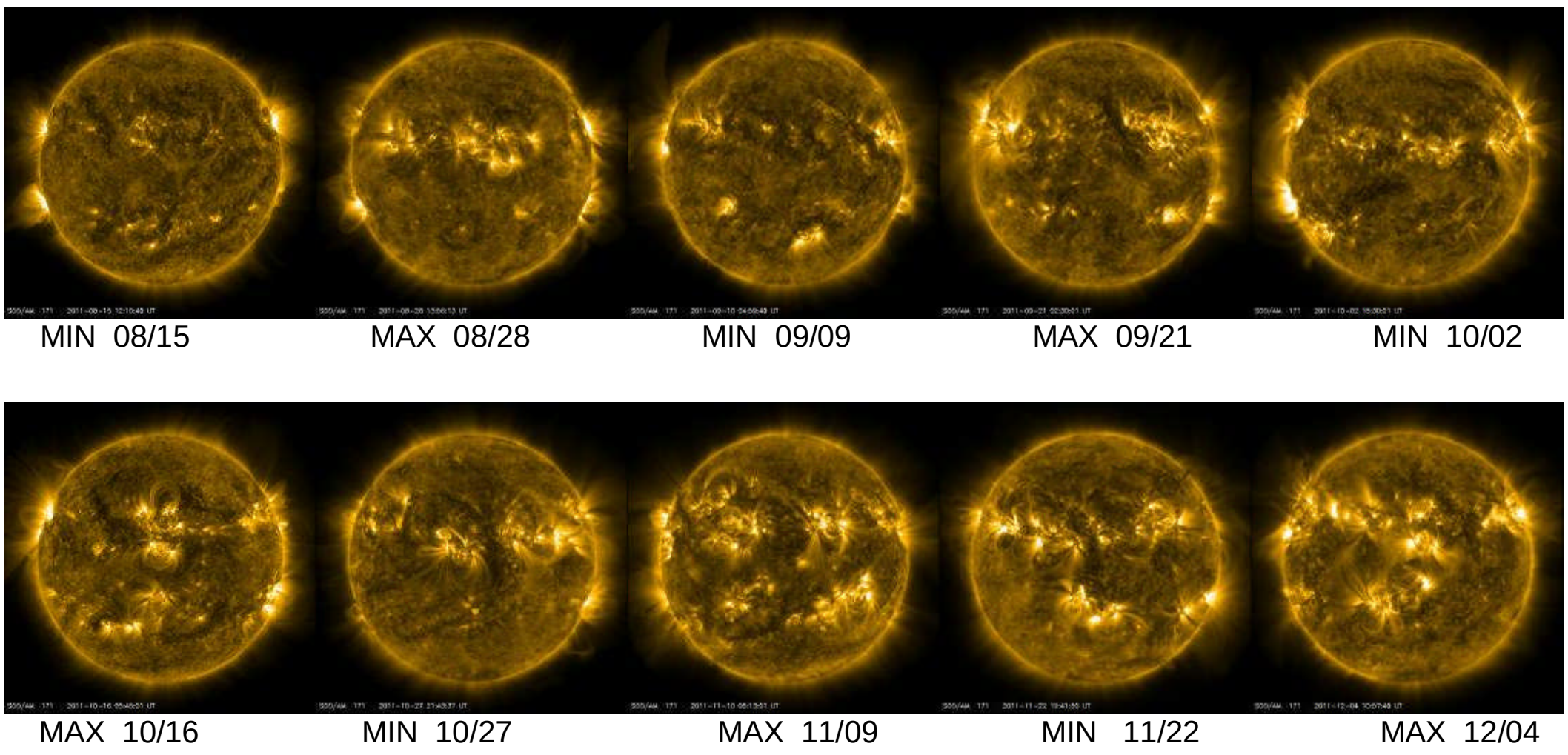

Figure 12. A sequence of images associated with the emergence of a region of solar activity on the Sun during the second half of 2011. Dates indicated are the dates of maxima and minima of $F_{10.7}$ solar radio flux as indicated in Figure 11.

It is clear from the images that, in the early stages of the SGE, sunspots emerged on only one side of the Sun. As that side of the Sun rotated into view of Earth the $\Delta_{27}F_{10.7}$ variation increased to a maximum then decreased to a minimum as the sunspots rotated behind the Sun. As the SGE progressed a new sunspot group began to emerge on the opposite side of the Sun. As a result, during the last solar rotation shown in Figure 12, it becomes difficult to distinguish, visually, between maximum and minimum of activity on the basis of sunspots. Referring to Figure 11 it appears that during the early phase of the SGE, when there are sunspots on only one side of the Sun, the $T_{MAX}$ variation moves into phase with the $\Delta_{27}F_{10.7}$ variation. However, as the active region becomes more symmetrically distributed over the Sun, or as a second SGE begins, the $\Delta_{27}T_{MAX}$ variation begins to drift towards out-of-phase with $\Delta_{27}F_{10.7}$ the variation. The physics associated with SGE is complex and not well understood, Charbonneau (2010).

**3.6 Extending the study backward in time.**

Solar activity has been studied extensively and accurately only during the satellite era from about 1979. The continuous daily $F_{10.7}$ index extends back to 1947. However, daily sunspot area (SSA) extends back to 1870 and daily sunspot number extends back to 1818. Australian temperature records extend back to 1853 (Melbourne Regional, $T_{MEL}$) while Central England temperature is available back to 1659. Thus there is considerable scope for extending this study of ~27 day period solar activity and temperature response over a wider time range. When the band pass method is applied to the raw data to obtain the $_{27}T_{MEL}$ and $_{27}$SSA(Nth) components between 1870 and 2013 the record can be visually scanned to detect significant variation in SSA and temperature. For example in 1928, Figure 13A, two SGE can be distinguished, one at the start of 1928 and one at the end of 1928, clearly associated with anti-phase and in-phase surface temperature variations respectively. In Figure 13B the $_{27}T_{MEL}$ variation is superimposed on the raw $T_{MEL}$ variation by addition of $S_{33}T_{MEL}$ and $_{27}T_{MEL}$. The 6$^{o}$C amplitude, ~27 day period, temperature variation at the start of 1928 is unequivocal evidence that ~27 day solar activity can have a very significant influence on daily surface temperature in Australia.

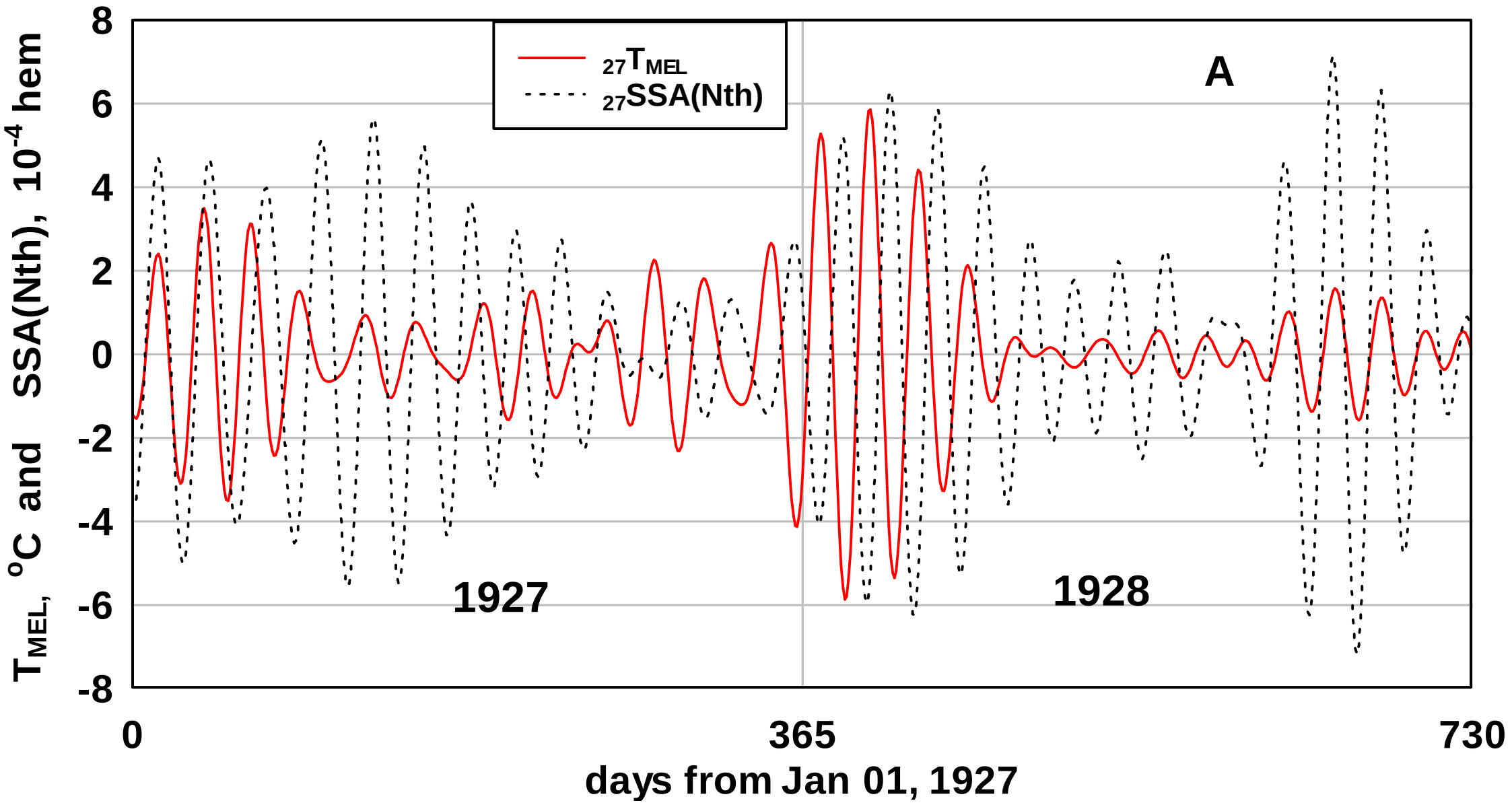

Figure 13. (A) Compares band pass filtered components of daily sunspot area, SSA(Nth), and daily maximum Melbourne temperature, $T_{MEL}$, for years 1927 and 1928.

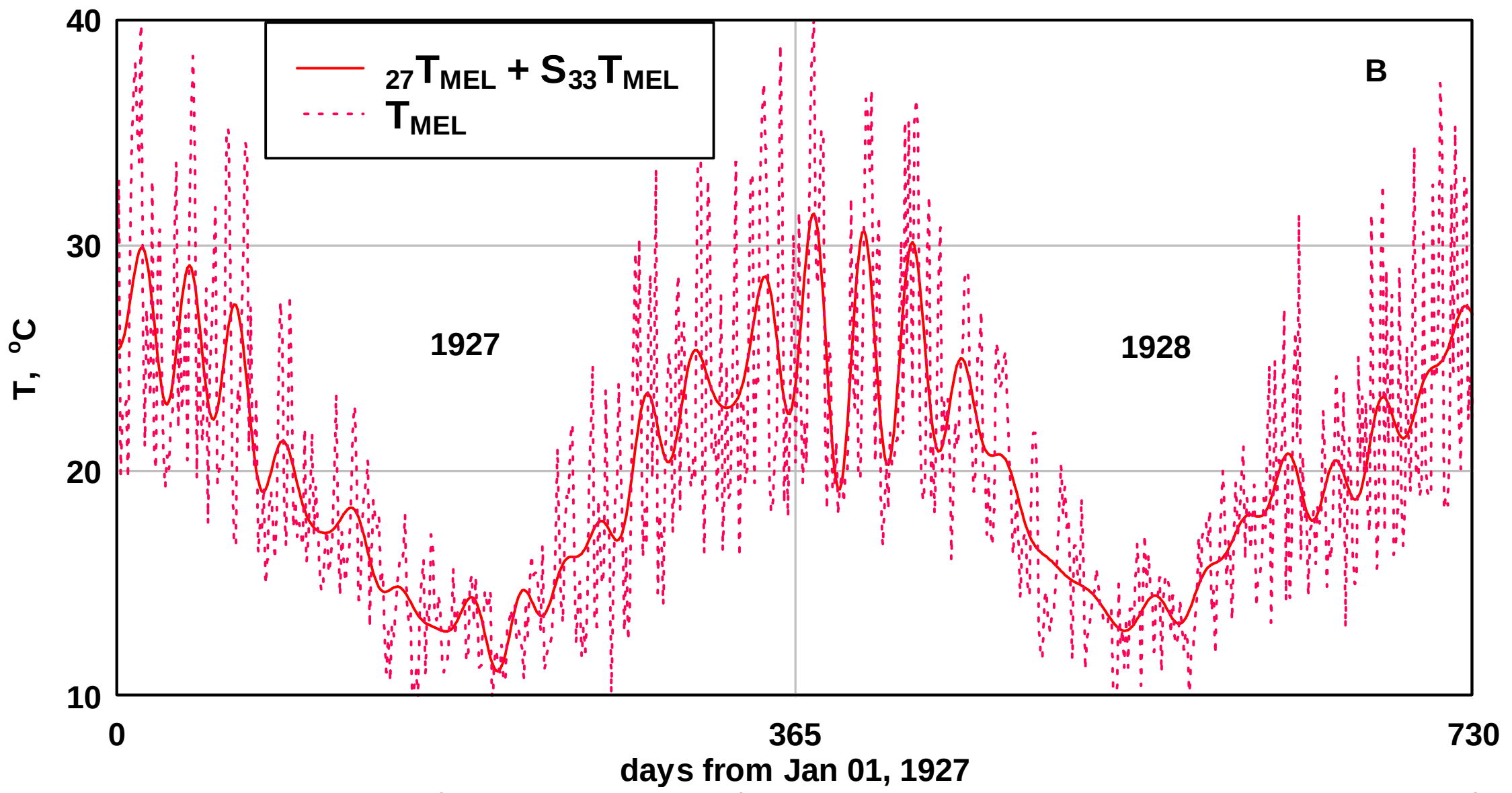

Figure 13 (B) The band pass filtered component of $T_{MEL}$ superimposed on the raw temperature data for Melbourne, 1927 and 1928.

## 4. Discussion.

We report a ~ 27 day period variation of Australian surface temperature that is correlated with the variation of solar activity and, therefore, correlated with the emergence and evolution of sunspot groups or active regions on the Sun. As SGE usually occur over a limited solar longitude range the solar activity, measured, on Earth, by proxies such as $F_{10.7}$ solar radio flux and SSA is modulated by the ~27 day period solar rotation and, as this paper demonstrates, this results in a ~ 27 day period temperature variation in Australia. The amplitude of the ~27 day temperature variation is significant, sometimes as high as 6$^o$C, and comparable with the seasonal temperature amplitude of about 6$^o$C and the amplitude of the shorter period synoptic variations, 3$^o$C to 8$^o$C, depending on season and location. We find that this temperature response to solar activity occurs, correlated with SGE, over most of the Australian continent and most strongly in the centre of the continent.

This effect has not previously been reported. Also, the amplitude of the effect, as high as 6$^o$C at times is, at first sight, energetically improbable. In comparison, the amplitude of the ~11 year cycle variation in global surface temperature is about 0.05$^o$C, Lean and Rind (2008). This paper focused on providing observations that clearly establish the association, in both amplitude and phase, of the ~ 27 day period surface temperature variation with the ~27 day variation in $F_{10.7}$ flux or SSA. In this we were guided by the paper of Volland and Schaefer (1979) that critiqued early work on Sun – weather relations e.g. King et al (1977), Wilcox et al, (1974), for failure to establish the coherence of Sun activity and weather variables. In this paper coherence was established by presenting observations in several annual periods when the amplitude of $\Delta_{27}T_{MAX}$ variation clearly varied in step with the amplitude of $\Delta F_{10.7}$ or $\Delta_{27}$SSA variation. The effect of the $\Delta_{27}T_{MAX}$ variation being mainly in-phase or out-of-phase with $\Delta_{27}F_{10.7}$ or $\Delta_{27}$SSA variation was also evident in the same annual periods. In addition spatial correlation of the $\Delta_{27}T_{MAX}$ variation and the $\Delta F_{10.7}$ variation was established for widely spaced locations in Australia for the case of one well defined SGE in 1975.

Outstanding questions arising from this work are (a), the high response of surface temperature over the Australian continent to solar activity associated with SGE; (b), why this response is limited to the continental area; (c), why the temperature response during a SGE can be either in phase or out of phase with the solar activity and (d), how to assess the temperature response when solar activity is intense and SGE overlap temporally at solar maximum. Clearly the answer to these questions depends on identifying the physical mechanism connecting ~27 day solar activity and surface temperature. Here we focus on the standing planetary wave mechanism developed by King et al (1977) and Volland (1979) as some of the observations in this paper appear to be consistent with this mechanism. However, the influence of other mechanisms e.g. Tinsley et al (2007), Haigh (2003) cannot be excluded.

King et al (1977) found, by superposed epoch analysis, a significant ~27 day period variation in the height of the 500 mbar pressure level at $70^{o}$N. They were able to demonstrate a ~27 day period standing wave pattern of wave number m = 1 and m = 2 type with largest antinodes at $140^{o}$E and $340^{o}$E. They found opposing maximum variations in surface pressure occurred at about $30^{o}$– $40^{o}$N and $70^{o}$N. Following the King et al (1977) report Volland (1979) showed theoretically that a ~ 27 day period variation of 0.1% in the solar radiation due to “long-lived co-rotating active longitudes on the Sun”, (called SGE in this paper), would preferentially heat continental land areas and this heating process would resonantly generate planetary waves of zonal wave number m = 2. The planetary waves, with meridional wave numbers n = 4, 5, 6 and 7 propagate eastward and westward and would generate standing longitudinal waves similar, in meridional distribution (Figure 14), to the standing waves observed by King et al (1977). However, it is worth noting that the predicted wave amplitude, about 0.4 mbar, is much less than the higher wave amplitudes observed by King et al (1977), about 12 mbar. So some other effect, probably cloud formation, must also be involved. Volland (1979) showed that any combination of the waves would result in a standing wave and a larger eastward propagating wave both with a meridional structure similar to the n = 6 wave in Figure 14.

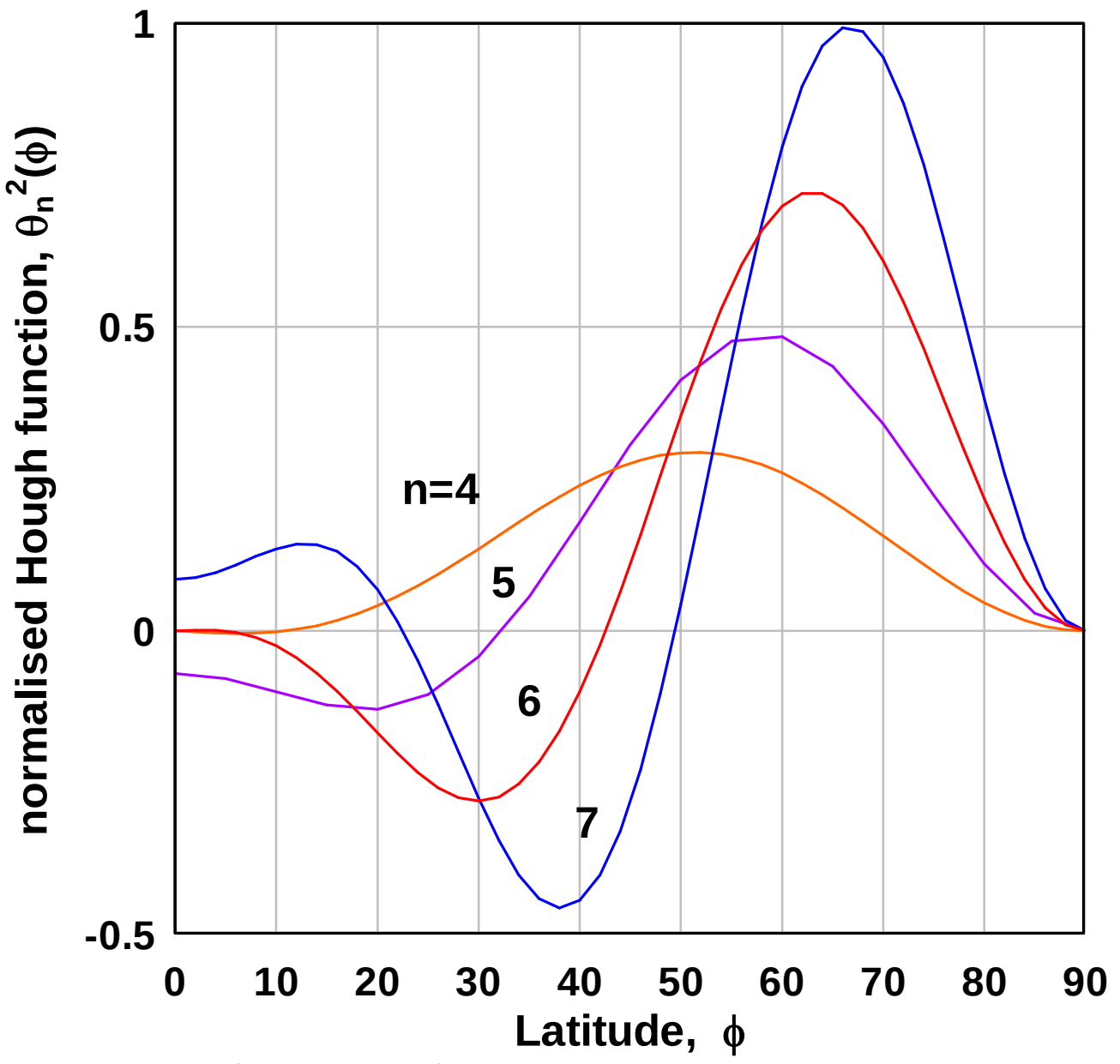

Figure 14. Normalised Hough functions of zonal number m = 2 and meridional wave number, n, according to equation (7), Volland (1979).

The King and Volland results are relevant to the observations in this paper as variation in height of the 500 mbar pressure level is directly related to variation in surface air temperature. The spatial distribution of the ~27 day response to an isolated SGE in 1975, Figure 9 and Figure 10, can be compared to the spatial distribution of a (m = 2, n = 6) standing planetary wave. The latitudinal extent of the weaker anti-node, 10S to 45S, in Figure 14 corresponds closely to the observed latitudinal response of the ~27 day component of temperature during the lifetime of the SGE in 1975, Figure 9. The longitude of the peak of the larger zonal anti-node, 140E, observed by King et al (1977) corresponds closely to the longitude of central Australia (135E) where the observed distribution of the ~27 day temperature response reaches a peak, Figure 10. An m = 2 planetary wave has four antinodes in $360^o$. Thus with an anti-node at central Australia (24S, 135E) the corresponding antipodal anti-node in the Northern Hemisphere would be in Western Sahara (24N, 315E). This spatial distribution would be consistent with the concept (Volland 1979) of a ~27 day period heat input to the Earth's continents generating planetary waves. Thus the Volland mechanism would seem to provide a partial answer to question (a), regarding the size of the response and to question (b), regarding the spatial distribution of the response. In respect to question (c), regarding the variable phase lag, the mechanism is also useful. Volland (1979) pointed out that any combination of the waves shown in Figure 14 would lead to standing waves similar to the m = 2, n = 6 wave. As the phase of the response at any specific latitude varies with the mode number, n, it is evident that different combinations of the modal response to solar activity would lead to different phase lags. Similarly, if different modes were excited at different times the observed phase lag at a specific location would vary from in-phase to out-of-phase. For example, a mode 5 wave, Figure 14, would give in-phase variation of temperature with solar activity at latitude $40^o$ whereas a mode 6 wave would give an out-of-phase variation. In respect to question (d), regarding the temperature response at solar maximum when many SGE occur simultaneously, the Volland mechanism is not immediately helpful. The variation of solar activity at solar maximum is not simply periodic at ~27 days but a quasi-periodic variation at ~27 days with rapidly shifting phase. Thus a single ~27 day component is expected to be partly replaced by a spectrum of components in sidebands associated with the phase modulation. Further analysis of how the Volland mechanism might apply to this situation is outside the scope of the paper.

The history of the Volland mechanism is interesting. In the 1970's there was much interest in the effect of ~27 day period solar activity on the troposphere, e.g. Wilcox (1974), King (1977), leading to the theory of resonant excitation of planetary waves by continental heating, Volland (1979). However, Schaefer (1978), repeating the study by King et al (1977) and using epoch superposition at intervals quite different from 27 days, found similar results to King et al (1977). Furthermore, spectral analysis of the 500 mbar height record at each measurement location did not show clearly dominant peaks near a period of 27 days. Subsequently Volland and Schaefer (1979) reviewed the observations and concluded that it was an "accidental coincidence" that turbulent processes on the Sun and Earth generate persistent wave structures with similar lifetimes and periods. And, if one observes both systems over time intervals not very long compared with the lifetime of the turbulent cells, a false impression of a real physical correlation is obtained. Based on the observations in this paper we come to the opposite conclusion: Localised activity on the Sun is associated with a ~27 day period variation in surface temperature that is positively or negatively correlated with the ~27 day variation of the activity while the activity persists. Applying correlation or spectral analyses over time intervals very long compared with the lifetime of the localised activity gives a false impression that no significant correlation exists.